\documentclass[aps,pra,twocolumn,reprint,superscriptaddress]{revtex4-1}

\usepackage{amsmath,amsthm,amssymb,amsfonts,bbm}
\usepackage{hyperref}

\newtheorem{thm}{Theorem}

\newtheorem{lem}{Lemma}
\newtheorem{prop}{Proposition}
\newtheorem{defn}{Definition}
\theoremstyle{remark}

\usepackage[usenames,dvipsnames]{xcolor}
\hypersetup{colorlinks=true,
						linkcolor=MidnightBlue,        
						citecolor=MidnightBlue,        
						filecolor=magenta,      			 
						urlcolor=MidnightBlue          
}

\begin{document}
\title{Necessity of Eigenstate Thermalization}

\author{Giacomo De Palma}
\affiliation{NEST, Scuola Normale Superiore and Istituto Nanoscienze-CNR, I-56127 Pisa,
Italy.}
\affiliation{INFN, Pisa, Italy}

\author{Alessio Serafini}
\affiliation{Department of Physics \& Astronomy, University College London, Gower Street, London WC1E 6BT, United Kingdom}
\affiliation{NEST, Scuola Normale Superiore and Istituto Nanoscienze-CNR, I-56127 Pisa,
Italy.}

\author{Vittorio Giovannetti}
\affiliation{NEST, Scuola Normale Superiore and Istituto Nanoscienze-CNR, I-56127 Pisa,
Italy.}

\author{Marcus Cramer}
\affiliation{Institut f\"ur Theoretische Physik, Universit\"at Ulm, Germany}

\begin{abstract}
Under the Eigenstate Thermalization Hypothesis (ETH), quantum-quenched systems equilibrate towards canonical, thermal ensembles. While at first glance the ETH might seem a very strong hypothesis, we show that it is indeed not only sufficient but also necessary for thermalization. More specifically, we consider systems coupled to baths with well-defined macroscopic temperature and show that whenever all product states thermalize then the ETH must hold. Our result definitively settles the question of determining whether a quantum system has a thermal behaviour, reducing it to checking whether its Hamiltonian satisfies the ETH.
\end{abstract}

\maketitle

An ideal heat bath induces thermalization in the sense that, when a physical system is coupled to it, its state will evolve to a well-defined infinite-time limit which depends only on macroscopic parameters of the bath -- such as its temperature or energy -- and not on any details of the initial state of the system, the bath, or the system-bath interaction.
It is a well-established empirical fact that both classical and quantum systems with a very large number of degrees of freedom exhibit these ideal-bath properties when weakly coupled to much smaller systems, with their temperature being a smooth function of their energy alone.
Yet, rigorous derivations relaying such a `generic' behaviour to fundamental dynamical laws
seem to require rather sophisticated, and arguably very specific and technical, hypotheses.
Then, understanding the mechanisms lying behind the thermalization of a quantum system has become a hot-debated topic in physics.
The apparent incongruence between the ubiquity of thermalization and the specificity of the hypotheses that seem to imply it has spurred substantial research~\cite{deutsch1991quantum,srednicki1994chaos,tasaki1998quantum,calabrese2006time,cazalilla2006effect,rigol2007relaxation,reimann2007typicality,cramer2008exact,rigol2008thermalization,reimann2008foundation,linden2009quantum,rigol2009breakdown,rigol2012alternatives,reimann2010canonical,cho2010emergence,gogolin2011absence,riera2012thermalization,mueller2013thermalization,gogolin2015equilibration,polkovnikov2011colloquium,cazalilla2011one,bloch2008many,eisert2015quantum,deffner2015ten,jarzynski2015diverse,ponte2015many,steinigeweg2014pushing,genway2013dynamics,caux2013time,cassidy2011generalized},
analyzing the dynamical conditions under which a large quantum system behaves as an ideal heat bath and induces thermalization.
Prominent among them is the Eigenstate Thermalization Hypothesis (ETH), which
may be formulated by stating that the partial traces of the eigenstates
of the global Hamiltonian of the bath and the coupled system (including the interaction terms)
are smooth functions of the energy.

It is well known that the ETH is sufficient for thermalization if the initial state has a sufficiently sharp distribution
in energy \cite{deutsch1991quantum,gogolin2015equilibration},
and a lot of effort has then been dedicated in checking whether specific quantum systems satisfy the ETH, with both analytical and numerical computations~\cite{rigol2008thermalization,ponte2015many,steinigeweg2014pushing,genway2013dynamics,caux2013time,cassidy2011generalized,rigol2009breakdown}.

The converse question, however, of whether the ETH is also necessary for thermalization, i.e. whether there exist quantum systems not fulfilling the ETH but nonetheless exibiting thermal behavior, is not settled yet, and alternatives to the ETH have been proposed~\cite{rigol2012alternatives}.
An answer to this question
has been hinted at, although not proven, in the literature on the subject (see, e.g., the very recent
survey \cite{gogolin2015equilibration}, to which the reader is also referred for a comprehensive overview of the context).
The goal of the present work is to clarify this subtle and somewhat elusive point by providing, for the first time to our knowledge,
a proof that the very definition of ideal bath actually implies the ETH.
Our result then definitively settles the question of determining whether a quantum system has a thermal behaviour, reducing it to checking whether its Hamiltonian satisfies the ETH: if the ETH is satisfied, the system always thermalizes, while if it is not satisfied, there certainly exists some reasonable physical initial state not leading to thermalization.

Before starting with the actual proof,
we find it mandatory to state preliminary, rigorous definitions of
thermalization, of an ideal bath and of the ETH itself.
We will then reconsider the role of the ETH as a sufficient condition for thermalization on the basis of our definitions,
and then proceed to present of our main finding, that the ETH is also necessary for thermalization.
Complete proofs of the lemmata needed in the paper may be found in Appendix \ref{app}.

\section{Thermalization and ideal baths}
Consider a system $S$ coupled to a heat bath $B$, with Hilbert spaces $\mathcal{H}_S$ and $\mathcal{H}_B$ of dimension $d_S$ and $d_B$, respectively.
For convenience, we describe the total Hamiltonian as $\hat{H}=\hat{H}_C+ \hat{H}_B$, composed of a free term $\hat{H}_B$ associated with the bath's inner dynamics, and a term $\hat{H}_C$ that includes
both the free component associated with $S$ and the system-bath coupling component. We only require the norm $\|\hat{H}_C\|$ to be bounded independently of the dimension $d_B$ of the bath~\cite{norms}.
Let then the global system start in some state $\hat{\rho}$. At time $t$ it will evolve into the density matrix $\hat{\rho}(t)=e^{-i\hat{H}t}\,\hat{\rho}\,e^{i\hat{H}t}$ whose time-averaged counterpart is
the diagonal part of $\hat{\rho}$ in the energy eigenbasis, $\Phi\left(\hat{\rho}\right)=\sum_n p_n\,|n\rangle\langle n|$,
assuming the spectrum of $\hat{H}$ to be non-degenerate for simplicity. Here, $\Phi$ denotes the time-averaging map and
$p_n=\langle n|\hat{\rho}|n\rangle$ is the probability that the global system has energy $E_n$~\cite{linden2009quantum}.
The time-averaged reduced state of the system $S$ is then obtained by taking the partial trace of
$\Phi\left(\hat{\rho}\right)$ over the bath degrees of freedom,
\begin{equation}\label{Phi}
\Phi_S\left(\hat{\rho}\right)\equiv\mathrm{Tr}_B\Phi\left(\hat{\rho}\right)=\sum_n p_n\;\hat{\tau}_n\;,
\end{equation}
where $\hat{\tau}_n\equiv\mathrm{Tr}_B|n\rangle\langle n|$ is the partial trace of the eigenstate $|n\rangle$.
In this context, thermalization is said to occur when the density matrices $\Phi_S\left(\hat{\rho}\right)$ exhibit a functional dependence only on those properties of the initial states $\hat{\rho}$
which are directly associated with the bath, as the initial properties of $S$ are washed away by the time-average and partial trace operations.

A key point in the study of such processes
is the choice of the set which identifies the initial states $\hat{\rho}$ of the joint system under which thermalization is assumed to occur:
too broad a set being typically too restrictive to describe realistic configurations, too narrow a set leading instead to trivial results.
In many cases of physical interest, one would know the value of only some macroscopic observables of the bath, such as the energy, so a common hypothesis is to impose thermalization when the bath is in the mixed state that maximizes the von Neumann entropy among all the states with given expectation values of the known observables \cite{reimann2010canonical}.
A weakness of this approach is that it does not account for situations where the bath is prepared in a pure state.
Another approach based on typicality has then been proposed.
In Ref. \cite{linden2009quantum}, the initial state of the bath is a pure state chosen randomly according to the Haar measure on the subspace of the bath Hilbert space compatible with the values of the known macroscopic observables.
The reduced system equilibrium state is then proven to be close, with very high probability,
to the equilibrium state resulting from choosing as initial state of the bath the normalized projector over the considered subspace.
A more refined choice would be to modify the notion of typicality by adopting
probability measures that reflect the complexity of the state preparation.
Indeed, the quantum pure states that are more easily built and comparatively stable are the ground states of local Hamiltonians,
so that one may restrict to the uniform measure on the states satisfying the area law \cite{eisert2010colloquium,garnerone2010typicality,garnerone2010statistical},
or introduce a measure arising from applying a local random quantum circuit to a completely factorized initial state \cite{hamma2012quantum,hamma2012ensemble}.
However, these probability measures are much more complicated than the uniform one on the whole Hilbert space,
and the computations may not be feasible.

Besides, asking whether there exist initial states of the bath not leading to thermalization of the system is a legitimate question,
to which these approaches based on typicality do not answer.
In this paper, we want to address precisely this question.
Our definition of thermalization is therefore:

\begin{defn}[Thermalization for initial product states]\label{defth}
We say that a subspace $\mathcal{H}_B^{\mathrm{eq}}$ of the bath Hilbert space induces thermalization of the system to a state $\hat{\omega}$ with precision $\epsilon$ if for any initial \emph{product} global state supported on $\mathcal{H}_S\otimes\mathcal{H}_B^{\mathrm{eq}}$ the equilibrium reduced state of the system is close to $\hat{\omega}$. That is,
$\mathcal{H}_B^{\mathrm{eq}}$ is such that \cite{norms}
\begin{equation}\label{thomega}
\left\|\Phi_S\left(\hat{\rho}\right)-\hat{\omega}\right\|_1\leq\epsilon
\end{equation}
for all $\hat{\rho}=\hat{\rho}_S\otimes\hat{\rho}_B$ with $\mathrm{Supp}\,\hat{\rho}_B\subset\mathcal{H}_B^{\mathrm{eq}}$.
\end{defn}
To discuss the connection between ETH and thermalization we shall further restrict the analysis to subspaces $\mathcal{H}_B^{\mathrm{eq}}$ corresponding to micro-canonical energy shells  $\mathcal{H}_B(E,\Delta_B)$ of the bath free Hamiltonian, i.e., to subspaces
 spanned by those eigenvectors of $\hat{H}_B$ with eigenvalues in the interval $[E-\Delta_B,E+\Delta_B]$.
In this context the  associated equilibrium reduced state $\hat{\omega}$ entering Eq.~(\ref{thomega}) is  assumed to depend upon $\mathcal{H}_B(E,\Delta_B)$
only via a smooth function $\beta(E)$ of $E$, which effectively defines the inverse temperature $1/T(E)=k\beta(E)$ of the bath, $k$ being the Boltzmann's constant. Notice that $\hat{\omega}(\beta(E))$ and $\beta(E)$ are otherwise arbitrary~\cite{GIBBS}.
Of course, a necessary condition for this to happen  is to have  the width $\Delta_B$ much smaller than the scale over which the mapping $E\mapsto \hat{\omega}(\beta(E))$ varies appreciably.
More precisely, with $C \equiv dE/dT>0$ the bath's heat capacity, we must have that $\hat{\omega}(\beta)$ does not appreciably change for
variations of $\beta$ on the order
 $\delta\beta\approx\Delta_B|d\beta/dE|= k\beta^2\Delta_B/C$.
Considering that the largest energy scale that can be associated with the system alone is the operator norm $\|\hat{H}_C\|$, we can conclude that
 thermalization with precision $\epsilon$ is reasonable if $\|\hat{H}_C\|\delta\beta\leq\epsilon$, i.e. if
\begin{equation}\label{condepsilon}
k\,{\beta(E)}^2\,\Delta_B\,\|\hat{H}_C\|\leq \epsilon\, C(\beta(E))\;.
\end{equation}
We are then led to define an ideal heat bath as follows.
\begin{defn}[Ideal heat bath]\label{bath}
We say that a bath is ideal in the energy range $\mathcal{E}_B$~\cite{NOTA1} with energy-dependent inverse temperature $\beta(E)$ if, for any
$\Delta_B$ and $\epsilon$ satisfying \eqref{condepsilon}
and for any $E\in\mathcal{E}_B$, the micro-canonical shell $\mathcal{H}_B(E,\Delta_B)$ induces thermalization to the state $\hat{\omega}(\beta(E))$ with precision $\epsilon$ in the sense of Definition \ref{defth}.
\end{defn}

\section{ETH implies thermalization}
The ETH roughly states that, given two eigenvalues $E_n$ and $E_m$ of the global Hamiltonian $\hat{H}$ which are close, the  associated reduced density matrices $\hat{\tau}_n$ and $\hat{\tau}_m$ defined in Eq.~(\ref{Phi})
must also be close,  i.e. that $\hat{\tau}_n$ is a ``sufficiently continuous'' function of the energy of the joint system.
More precisely,
our working definition is the following:
\begin{defn}[ETH]\label{ETH}
We say that a Hamiltonian $\hat{H}=\sum_nE_n|n\rangle\langle n|$ fulfils the ETH in the region of the spectrum $\mathcal{E}$~\cite{NOTA1}
 on a scale $\Delta$ with precision $\epsilon_{eth}$  if
all $E_n,\,E_m\in\mathcal{E}$ with $\left|E_m-E_n\right|\le 2\Delta$ fulfil  $\left\|\hat{\tau}_m-\hat{\tau}_n\right\|_1\le \epsilon_{eth}$.
\end{defn}
It is worth observing that
the usual formulation of the ETH \cite{deutsch1991quantum,gogolin2015equilibration} does not split the global system into system and bath. Instead, it identifies a class of relevant macroscopic observables $\mathcal{A}$, and states that for any $\hat{A}\in\mathcal{A}$ the diagonal matrix elements in the energy eigenbasis $\langle n|\hat{A}|n\rangle$ depend ``sufficiently continuously'' on the energy. Upon choosing as $\mathcal{A}$ the set of self-adjoint operators acting on the system alone, our definition is equivalent. Indeed, for any $\hat{A}=\hat{A}_S\otimes\hat{\mathbb{I}}_B$ we have $\langle n|\hat{A}|n\rangle=\mathrm{Tr}_S(\hat{A}_S\,\hat{\tau}_n)$, which are sufficiently continuous functions of the energy for any $\hat{A}_S$ if and only if $\hat{\tau}_n$ is.

It is well-established that if the ETH holds for any initial global state with a sharp enough energy distribution, then the time average of the reduced state of the system is a smooth function of its average global energy alone \cite{gogolin2015equilibration}, i.e. different initial global states lead to nearly the same equilibrium reduced state for the system if their average energies are close and their energy distribution is sufficiently sharp.
Moreover, this equilibrium state is close to the one associated with a micro-canonical global state.
To make our treatment self-contained, and better emphasise the importance of the ETH in the study of thermalization,
let us state here precisely our version of this implication in terms of the definitions introduced above (see Section \ref{proofP1} for a proof).
\begin{prop}[ETH implies micro-canonical thermalization]\label{ETH->th}
Let $\hat{H}$ fulfil the ETH in $\mathcal{E}$ on a scale $\Delta$ with precision $\epsilon_{eth}$.
Let $\hat{P}$ be the projector onto the energy shell $\mathcal{H}(E,\Delta)$ of the total Hamiltonian, so onto the subspace spanned by those eigenvectors of $\hat{H}$ that have eigenvalues in the interval $[E-\Delta,E+\Delta]$, which is assumed to be contained in $\mathcal{E}$.
Then, for any initial state $\hat{\rho}$ peaked around the energy $E$ in the sense $\mathrm{Tr}[\hat{\rho}(\hat{\mathbb{I}}-\hat{P})]\leq\epsilon_{eth}$,
the time-averaged reduced state $\Phi_S\left(\hat{\rho}\right)$ of Eq.~(\ref{Phi}) is close to the reduced micro-canonical state associated to $\mathcal{H}(E,\Delta)$,
\begin{equation}\label{omegaE}
\left\|\Phi_S(\hat{\rho})-\mathrm{Tr}_B(\hat{P})/\mathrm{Tr}(\hat{P})\right\|_1\le 3\epsilon_{eth}\;.
\end{equation}
\end{prop}
Let us stress that this proposition does \emph{not} assume $\hat{\rho}$ to be a product or separable state, i.e., the ETH implies thermalization even if the system and bath are initially entangled.
The link with Definitions \ref{defth} and \ref{bath} is then provided by Lemma~\ref{rhoQ} of Appendix \ref{app}: If $\hat{\rho}$ is a state supported on $\mathcal{H}_S\otimes\mathcal{H}_B(E,\Delta_B)$ then $\mathrm{Tr}[\hat{\rho}(\hat{\mathbb{I}}-\hat{P})]\leq\epsilon_{eth}$ and Eq.~\eqref{omegaE} follows from the ETH on a scale $\Delta=(\|\hat{H}_C\|+\Delta_B)/\sqrt{\epsilon_{eth}}$. Further, for conditions under which the micro-canonical state may be replaced by the canonical state, see, e.g., Refs.~\cite{riera2012thermalization,mueller2013thermalization,popescu2006entanglement,goldstein2006canonical,brandao2015equivalence} and references therein.

\section{Thermalization implies ETH}
Proposition~\ref{ETH->th} seems to imply that the ETH is too strong a hypothesis and that weaker assumptions might be sufficient to
justify thermalization. It turns out that this is not true. Indeed,
we shall prove  that the ETH must hold for any ideal heat bath satisfying Definition \ref{bath}.
First off, we show that if a subspace of the bath $\mathcal{H}_B^{\mathrm{eq}}$ induces thermalization to a state $\hat{\omega}$ for any initial product state as per Definition \ref{defth}, the property extends to the entangled initial states up to an overhead which is linear in the system dimension.
Our argument relies on the observation that the entanglement of the eigenstates $|n\rangle$ is limited by the system dimension $d_S$, and cannot grow arbitrarily even when the bath dimension is large.
Note that this result is similar in spirit to the main finding of Ref. \cite{gogolin2011absence}, where thermalization is disproved in certain non-integrable systems by establishing an upper bound on the average system-bath entanglement over random initial bath states.
\begin{lem}\label{therm}
Let $\mathcal{H}_B^{\mathrm{eq}}$ be a subspace of the bath Hilbert space that induces thermalization to a state $\hat{\omega}$ with precision $\epsilon$ in the sense of Definition \ref{defth}.
Then $\mathcal{H}_B^{\mathrm{eq}}$ induces thermalization also on the entangled initial states with precision $4d_S\epsilon$, i.e.
\begin{equation}\label{thm1}
\left\|\Phi_S\left(\hat{\rho}\right)-\hat{\omega}\right\|_1\leq4d_S\epsilon
\end{equation}
for all $\hat{\rho}$ with support contained in $\mathcal{H}_S\otimes\mathcal{H}_B^{\mathrm{eq}}$.
\end{lem}
By virtue of this Lemma,
the equilibration to some fixed state $\hat{\omega}$ of all initial product states in $\mathcal{H}_S\otimes\mathcal{H}_B^{\mathrm{eq}}$ extends to {\it all} initial states in this subspace.
Then, if an eigenstate $|n\rangle$ of the Hamiltonian is almost contained in the same subspace,
the resulting time-averaged reduced state of the system $\Phi_S(|n\rangle\langle n|)$ is also close to $\hat{\omega}$.
However, if we initialize the global system in an eigenstate of the Hamiltonian, it obviously remains there forever,
\begin{equation}\label{Phin}
\Phi_S(|n\rangle\langle n|)=\hat{\tau}_n\;.
\end{equation}
Combining this with the fact that the trace norm is contracting under completely positive trace-preserving maps \cite{perez2006contractivity}, we have under the assumptions of Lemma \ref{therm} that (see Section \ref{main proof} for details)
\begin{equation}
\label{old_lemma_2}
\left\|\hat{\tau}_n-\hat{\omega}\right\|_1
\le 4d_S\epsilon+2\sqrt{\langle n|\hat{Q}|n\rangle},
\end{equation}
where $\hat{Q}$ is the projector onto the subspace orthogonal to $\mathcal{H}_S\otimes\mathcal{H}_B^{\mathrm{eq}}$. It remains to bound $\langle n|\hat{Q}|n\rangle$ for given $\mathcal{H}_B^{\text{eq}}=\mathcal{H}_B(E,\Delta_B)$, which we do in Section \ref{main proof}, to arrive at the statement that whenever $\mathcal{H}_B(E,\Delta_B)$ induces thermalization to $\hat{\omega}$ with precision $\epsilon$ then for all $n$ with $|E_n-E|\le \Delta_B/2$ we have
\begin{equation}
\left\|\hat{\tau}_n-\hat{\omega}\right\|_1\le \frac{8\|\hat{H}_C\|^2}{\Delta_B^2}+4d_S\epsilon,
\end{equation}
which implies our main result (see Section \ref{main proof} for details):
\begin{thm}[Thermalization implies ETH]\label{->ETH} Let the bath be ideal in the energy range $\mathcal{E}_B$ as in Definition~\ref{bath}. Let
\begin{equation}
\epsilon_{eth}=12\sup_{E\in\mathcal{E}_B}
\left(\frac{2\|\hat{H}_C\|^2d_S\,k\,{\beta(E)}^2}{C(\beta(E))}\right)^{2/3}.
\end{equation}
Then $\hat{H}$ fulfils the ETH in the region $\mathcal{E}_B$ on a scale
\begin{equation}
\Delta=2\sqrt{3}\frac{\|\hat{H}_C\|}{\sqrt{\epsilon_{eth}}}
\end{equation} with precision $\epsilon_{eth}$.
\end{thm}

Typically,
for any fixed inverse temperature $\beta$, the bath's heat capacity $C(\beta)$ is increasing in the size of the bath.
On the contrary, $\hat{H}_C$ has been chosen such that it remains bounded.
Then, for fixed $\beta$ and $d_S$, the error $\epsilon$ becomes arbitrarily small (and thus the width $\Delta$ arbitrarily large) as $d_B\rightarrow\infty$.

\section{Conclusions}
The eigenstate thermalization hypothesis has been central to much of the ongoing discussion
concerning the relaxation of open quantum systems to fixed equilibrium states.
Its role as a sufficient condition for thermalization, which we reviewed in Proposition \ref{ETH->th}, is well established
and has been repeatedly remarked in several past contributions.
By proving that, conversely, an ideal heat bath must necessarily interact with the system with
a Hamiltonian fulfilling the ETH we have, in a precise and rigorous sense, revealed the full role
such a condition has to play. This result rests on a definition of an ideal bath which is rigorous and yet
broad enough to encompass all practically relevant instances,
and hence sheds considerable light on the very general mechanisms that let open quantum systems thermalize.

\acknowledgments
We thank C. Gogolin and P. Zanardi for their useful comments.
GdP thanks F. Essler and A. De Luca for useful discussions.
AS acknowledges financial support from EPSRC through grant EP/K026267/1 as well as the warm hospitality of SNS Pisa. MC acknowledges the EU Integrated Project SIQS and the Alexander von Humboldt foundation for financial support.

\appendix
\begin{widetext}
\section{}\label{app}
Here we provide explicit proofs of the various lemmata and theorems  presented in the main text.

\subsection{Proof of Proposition \ref{ETH->th}}\label{proofP1}
Defining
\begin{equation}\label{defC}
\mathcal{C}\equiv\left\{n:|E_n-E|\le\Delta\right\}\subset\mathcal{E}\;,
\end{equation}
the partial trace of the micro-canonical shell can be written as
\begin{equation}
\frac{\mathrm{Tr}_B\hat{P}}{\mathrm{Tr}\,\hat{P}}=\frac{1}{|\mathcal{C}|}\sum_{n\in\mathcal{C}}\hat{\tau}_n\;.
\end{equation}
We have then
\begin{eqnarray}
\left\|\Phi_S\left(\hat{\rho}\right)-\frac{\mathrm{Tr}_B\hat{P}}{\mathrm{Tr}\,\hat{P}}\right\|_1&=& \frac{1}{|\mathcal{C}|}\left\|\sum_n\sum_{m\in\mathcal{C}}p_n\left(\hat{\tau}_n-\hat{\tau}_m\right)\right\|_1\leq \frac{1}{|\mathcal{C}|}\sum_n\sum_{m\in\mathcal{C}}p_n\left\|\hat{\tau}_n-\hat{\tau}_m\right\|_1\leq\nonumber\\ &\leq&\frac{1}{|\mathcal{C}|}\sum_{n\in\mathcal{C}}\sum_{m\in\mathcal{C}}p_n\left\|\hat{\tau}_n-\hat{\tau}_m\right\|_1+2\mathrm{Tr}(\hat{\rho}\,(\hat{\mathbb{I}}-\hat{P}))\;,
\end{eqnarray}
where $\mathrm{Tr}(\hat{\rho}\,(\hat{\mathbb{I}}-\hat{P}))\le \epsilon_{eth}$ and
from \eqref{defC}, for any $m,n\in\mathcal{C}$ we have $|E_n-E_m|\le 2\Delta$ and then $\left\|\hat{\tau}_n-\hat{\tau}_m\right\|\leq\epsilon_{eth}$.

\subsection{Lemma \ref{rhoQ}}

One arrives at the statement after Proposition~1 in the main text by applying the following lemma to $\hat{A}_1=\hat{H}$, $\hat{A}_2=\hat{\mathbb{I}}_S\otimes\hat{H}_B$, $\lambda=E$, $\Delta_1=\Delta=\frac{\|\hat{H}_C\|+\Delta_B}{\sqrt{\epsilon_{eth}}}$ and $\Delta_2=\Delta_B$.
\begin{lem}\label{rhoQ}
Consider two self-adjoint operators $\hat{A}_1$ and $\hat{A}_2$.
Let $\mathcal{H}_i(\lambda,\Delta_i)$ be the subspace identified by $\lambda-\Delta_i\leq\hat{A}_i\leq\lambda+\Delta_i$, for $i=1,2$.
Let $\hat{\rho}$ be a quantum state with support contained in $\mathcal{H}_2(\lambda,\Delta_2)$, and $\hat{Q}$ the projector onto the subspace orthogonal to $\mathcal{H}_1(\lambda,\Delta_1)$.
Then
\begin{equation}
\label{dd}
\mathrm{Tr}\bigl(\hat{\rho}\,\hat{Q}\bigr)\leq \left(\frac{\|\hat{A}_1-\hat{A}_2\|+\Delta_2}{\Delta_1}\right)^2\;.
\end{equation}
\begin{proof}
Consider first a pure state $\hat{\rho}=|\psi\rangle\langle\psi|$ and start from the identity
\begin{equation}
(\hat{A}_1-\lambda)|\psi\rangle=(\hat{A}_1-\hat{A}_2)|\psi\rangle+(\hat{A}_2-\lambda)|\psi\rangle\;.
\end{equation}
On one hand, the square norm of the left-hand-side is
\begin{equation}\label{LHS}
\|(\hat{A}_1-\lambda)|\psi\rangle\|^2=\langle\psi|(\hat{A}_1-\lambda)^2|\psi\rangle\geq\Delta_1^2\langle\psi|\hat{Q}|\psi\rangle\;.
\end{equation}
On the other hand, the norm of the right-hand-side satisfies
\begin{equation}\label{RHS}
\|(\hat{A}_1-\hat{A}_2)|\psi\rangle+(\hat{A}_2-\lambda)|\psi\rangle\|\leq \|\hat{A}_1-\hat{A}_2\|+\|(\hat{A}_2-\lambda)|\psi\rangle\|\leq \|\hat{A}_1-\hat{A}_2\|+\sqrt{\langle\psi|(\hat{A}_2-\lambda)^2|\psi\rangle}\leq \|\hat{A}_1-\hat{A}_2\|+\Delta_2\;.
\end{equation}
Putting together \eqref{LHS} and \eqref{RHS}, we have Eq.~(\ref{dd}) for pure states. For mixed states $\hat{\rho}=\sum_k p_k|\psi_k\rangle\langle\psi_k|$ with  $|\psi_k\rangle\in\mathcal{H}_2(\lambda,\Delta_2)$ we have $\mathrm{Tr}(\hat{\rho}\,\hat{Q})=\sum_kp_k\langle\psi_k|\hat{Q}|\psi_k\rangle$ and the assertion follows by applying the pure-state result to each term individually.
\end{proof}
\end{lem}

\subsection{Proof of Lemma~\ref{therm}}\label{proofL1}
Let $\hat{\rho}$ a state in the subspace $\mathcal{H}_S\otimes\mathcal{H}_B^{eq}$ and denote by $\hat{P}$ the projector on said subspace.
By the variational characterization of the trace norm we have
\begin{equation}
\frac{1}{2}\|\Phi_S\left(\hat{\rho}\right)-\hat{\omega}\|_1=\bigl|\text{Tr}\bigl[\hat{M}(\Phi_S\left(\hat{\rho}\right)-\hat{\omega})\bigr]\bigr|
=\bigl|\text{Tr}[\hat{\rho}\hat{X}_{\hat{M}}]\bigr|,\;\;\;\hat{X}_{\hat{M}}=\sum_n\text{Tr}[(\hat{\tau}_n-\hat{\omega})\hat{M}]\hat{P}|n\rangle\langle n|\hat{P}\;,
\end{equation}
for some $\hat{M}$ with $0\le \hat{M}\le\hat{\mathbb{I}}$. Assuming w.l.o.g. that $\hat{\varrho}=|\psi\rangle\langle\psi|$ (the mixed case follows by convexity), we may Schmidt-decompose $|\psi\rangle=\sum_{k=1}^{d_S}\psi_k|s_k\rangle| b_k\rangle$. Hence
\begin{equation}
\frac{1}{2}\|\Phi_S\left(|\psi\rangle\langle\psi|\right)-\hat{\omega}\|_1=\sum_{k,l}\psi_k\psi_l\langle s_k|\langle b_k|\hat{X}_{\hat{M}}|s_l\rangle|b_l\rangle
=:\sum_{k,l}\psi_k\psi_l\langle s_k|\hat{S}_{k,l}|s_l\rangle\;,
\end{equation}
where the operator $\hat{S}_{k,l}=\langle b_k|\hat{X}_{\hat{M}}|b_l\rangle$ acts on $\mathcal{H}_S$ and we have
by the triangle and Cauchy--Schwarz inequality
\begin{equation}
2|\langle s_k|\hat{S}_{k,l}|s_l\rangle|\le |\langle s_k|\bigl(\hat{S}_{k,l}+\hat{S}_{k,l}^\dagger\bigr)|s_l\rangle|+|\langle s_k|i\bigl(\hat{S}_{k,l}-\hat{S}^\dagger_{k,l}\bigr)|s_l\rangle|\le \|\hat{S}_{k,l}+\hat{S}_{k,l}^\dagger\|+\|i\bigl(\hat{S}_{k,l}-\hat{S}^\dagger_{k,l}\bigr)\|,
\end{equation}
where $\hat{S}_{k,l}+\hat{S}_{k,l}^\dagger$ and $i(\hat{S}_{k,l}-\hat{S}^\dagger_{k,l})$ are hermitian such that
\begin{eqnarray}
\|\hat{S}_{k,l}+\hat{S}_{k,l}^\dagger\|&=& \max_{\substack{|\psi\rangle\in\mathcal{H}_S\\ \langle\psi|\psi\rangle=1}}|\langle \psi|\bigl(\hat{S}_{k,l}+\hat{S}_{k,l}^\dagger\bigr)|\psi\rangle|
\le \max_{\substack{|\psi\rangle\in\mathcal{H}_S\\ \langle\psi|\psi\rangle=1}}|\langle \psi|\langle b_k|\hat{X}_{\hat{M}}|b_l\rangle|\psi\rangle|
+\max_{\substack{|\psi\rangle\in\mathcal{H}_S\\ \langle\psi|\psi\rangle=1}}|\langle \psi|\langle b_l|\hat{X}_{\hat{M}}|b_k\rangle|\psi\rangle|\le\nonumber\\
&\le& 2\max_{\substack{|\psi\rangle\in\mathcal{H}_S\\ \langle\psi|\psi\rangle=1}}
\max_{\substack{|\phi\rangle\in\mathcal{H}_B^{\text{eq}}\\ \langle\phi|\phi\rangle=1}}
|\langle \psi|\langle \phi|\hat{X}_{\hat{M}}|\phi\rangle|\psi\rangle|,
\end{eqnarray}
where we used the Cauchy-Schwarz inequality and the hermiticity of $\langle \psi|\hat{X}_{\hat{M}}|\psi\rangle$ to obtain the last line. The same upper bound holds for $|\langle s_k|i\bigl(\hat{S}_{k,l}-\hat{S}^\dagger_{k,l}\bigr)|s_l\rangle|$. Further,
\begin{equation}
|\langle \psi|\langle \phi|\hat{X}_{\hat{M}}|\phi\rangle|\psi\rangle|=\bigl|\text{Tr}\bigl[\hat{M}(\Phi_S\left(|\psi\rangle\langle\psi|\otimes|\phi\rangle\langle\phi|\right)-\hat{\omega})\bigr]\bigr|\le
\bigl\|\Phi_S\left(|\psi\rangle\langle\psi|\otimes|\phi\rangle\langle\phi|\right)-\hat{\omega}\bigr\|_1\le \epsilon
\end{equation}
such that
\begin{equation}
\|\Phi_S\left(|\psi\rangle\langle\psi|\right)-\hat{\omega}\|_1\le 4\epsilon\sum_{k,l}\psi_k\psi_l\le 4\epsilon d_S.
\end{equation}

\subsection{Proof of Theorem~\ref{->ETH}}
\label{main proof}
We first give the details of how to arrive at Eq.~\eqref{old_lemma_2}. Denote the projector onto $\mathcal{H}_S\otimes\mathcal{H}_B^{\mathrm{eq}}$ by $\hat{P}$. Inserting a zero and using the triangle inequality yields
\begin{equation}
\left\|\hat{\tau}_n-\hat{\omega}\right\|_1
\le \left\|\Phi_S\left(|n\rangle\langle n|-\frac{\hat{P}|n\rangle\langle n|\hat{P}}{\langle n|\hat{P}|n\rangle}\right)\right\|_1+\left\|\Phi_S\left(\frac{\hat{P}|n\rangle\langle n|\hat{P}}{\langle n|\hat{P}|n\rangle}\right)-\hat{\omega}\right\|_1\;.
\end{equation}
Making use of the contractivity of the trace norm for the first term and the assumptions of Lemma~\ref{therm} for the second term, we have
\begin{equation}
\left\|\hat{\tau}_n-\hat{\omega}\right\|_1
\le \left\||n\rangle\langle n|-\frac{\hat{P}|n\rangle\langle n|\hat{P}}{\langle n|\hat{P}|n\rangle}\right\|_1+4d_S\epsilon=2\sqrt{\langle n|\hat{Q}|n\rangle}+4d_S\epsilon\;,\label{aa}
\end{equation}
where in the second step we have derived the trace norm with an explicit computation of the eigenvalues.

Now let $\hat{H}_B=\sum_{k}e_k|k\rangle\langle k|$ and $\hat{Q}=\sum_{k\notin\mathcal{H}_B^{\text{eq}}}|k\rangle\langle k|$. Then
\begin{equation}
\min_{k\notin \mathcal{H}_B^{\text{eq}}}(e_k-E_n)^2\hat{Q}\le \sum_{k\notin\mathcal{H}_B^{\text{eq}}}(e_k-E_n)^2|k\rangle\langle k|
=\hat{Q}\sum_{k}(e_k-E_n)^2|k\rangle\langle k|
=\hat{Q}\bigl(\hat{H}_B-E_n\bigr)^2.
\end{equation}
Hence, by the Cauchy--Schwarz inequality
\begin{eqnarray}\label{bb}
\min_{k\notin \mathcal{H}_B^{\text{eq}}}(e_k-E_n)^2\langle n|\hat{Q}|n\rangle&\le& \sqrt{\langle n|\hat{Q}|n\rangle\langle n|\bigl(\hat{H}_B-E_n\bigr)^4|n\rangle}
=\sqrt{\langle n|\hat{Q}|n\rangle\langle n|\bigl(\hat{H}_B-\hat{H}\bigr)^4|n\rangle}\le\nonumber\\
&\le& \sqrt{\langle n|\hat{Q}|n\rangle}\|\hat{H}_B-\hat{H}\|^2 =\sqrt{\langle n|\hat{Q}|n\rangle}\|\hat{H}_C\|^2.
\end{eqnarray}
With $\mathcal{H}_B^{\text{eq}}=\mathcal{H}_B(E,\Delta_B)=\text{span}\{|k\rangle : |e_k-E|\le \Delta_B\}$, we have
\begin{equation}
\min_{k\notin \mathcal{H}_B^{\text{eq}}}(e_k-E_n)^2=\min_{k:|e_k-E|>\Delta_B}(e_k-E_n)^2\;,
\end{equation}
and, combining Eqs.~(\ref{aa},\ref{bb}), we have that if  $\mathcal{H}_B^{\text{eq}}=\mathcal{H}_B(E,\Delta_B)$ induces thermalization to a state $\hat{\omega}$ with precision $\epsilon$ then for all $n$ with $|E_n-E|\le \Delta_B/2$
\begin{equation}
\left\|\hat{\tau}_n-\hat{\omega}\right\|_1\le \frac{2\|\hat{H}_C\|^2}{\min_{|e-E|>\Delta_B}(e-E_n)^2}+4d_S\epsilon\le \frac{8\|\hat{H}_C\|^2}{\Delta_B^2}+4d_S\epsilon.
\end{equation}

If the bath is ideal in the energy range $\mathcal{E}_B$ with inverse temperature $\beta(E)$ then for any $\epsilon$, $\Delta_B$ with
\begin{equation}
\label{cc}
k\,{\beta(E)}^2\,\Delta_B\,\|\hat{H}_C\|\le \epsilon\, C(\beta(E))
\end{equation}
and any $E\in\mathcal{E}_B$ we have that $\mathcal{H}_B(E,\Delta_B)$ induces thermalization to the state $\hat{\omega}(\beta(E))$ with precision $\epsilon$. Hence, setting $\epsilon$ such that we have equality in Eq.~(\ref{cc})
 and letting $E\in\mathcal{E}_B$ and $n$ such that $|E_n-E|\le \Delta_B/2$, we have
\begin{equation}
\left\|\hat{\tau}_n-\hat{\omega}\right\|_1\le 4\|\hat{H}_C\|\left(\frac{2\|\hat{H}_C\|}{\Delta_B^2}+d_S\frac{k\,{\beta(E)}^2\,\Delta_B\,}{C(\beta(E))}\right),
\end{equation}
which is minimized by
\begin{equation}
\Delta_B^3=\frac{4\|\hat{H}_C\|C(\beta(E))}{d_Sk\,{\beta(E)}^2}\;.
\end{equation}
Hence, if the bath is ideal in the energy range $\mathcal{E}_B$ then for any $E_n,E_m\in\mathcal{E}_B$ with (we set $E=(E_n+E_m)/2$)
\begin{equation}
|E_n/2-E_m/2|=|E_{n/m}-E|\le  \Delta_B/2:=\Delta=2\|\hat{H}_C\|\sqrt{\frac{3}{\epsilon_{eth}}}\;,
\end{equation}
 we have
\begin{equation}
\left\|\hat{\tau}_{n/m}-\hat{\omega}\right\|_1\le 24\|\hat{H}_C\|^{2}/\Delta_B^{2}=6\left(\frac{2\|\hat{H}_C\|^2d_Sk\,{\beta(E)}^2}{C(\beta(E))}\right)^{2/3}=:\epsilon_{eth}/2\;,
\end{equation}
and finally
\begin{equation}
\|\hat{\tau}_m-\hat{\tau}_n\|_1\le \left\|\hat{\tau}_m-\hat{\omega}\right\|_1+\left\|\hat{\omega}-\hat{\tau}_n\right\|_1\le\epsilon_{eth}\;.
\end{equation}
\end{widetext}


\begin{thebibliography}{99}
\bibitem{norms} Throughout the paper, we denote by $\|\cdot\|$ and $\|\cdot\|_1$ the operator norm (so the largest singular value) and trace norm (so the sum of the singular values) of $\cdot\;$, respectively.
\bibitem{GIBBS}
In thermodynamics, the inverse temperature $\beta(E)$ is related to the density of energy levels
of $\hat{H}_B$ around $E$, $\Omega(E)$, by
$\beta(E)= \partial_{E}\ln\Omega$, while the density matrices  $\hat{\omega}(\beta)$ are identified with
the Gibbs states associated with the system Hamiltonian $\hat{H}_S$, i.e. $\hat{\omega}(\beta)=e^{-\beta\hat{H}_S}/\mathrm{Tr}(e^{-\beta\hat{H}_S})$.
However, both these assumptions are not necessary to prove our results, and we shall not make them here.
\end{thebibliography}
\end{document}